\documentclass[12pt]{article}
\usepackage{color}
\usepackage{bbm}
\usepackage{graphicx}
\usepackage{amsmath}
\usepackage{amsfonts}
\usepackage{amssymb}
\usepackage{fullpage}
\usepackage{enumerate}

\newcommand{\vp}{\varphi}

\newcommand{\Beq}{\begin{equation}\begin{aligned}}
\newcommand{\Eeq}{\end{aligned}\end{equation}}

\newcommand{\fF}{\mathcal{F}}

\newcommand{\te}{ {\tilde{\epsilon}}}

\newcommand{\trho}{ {\tilde{\rho}}}
\newcommand{\mE}{ {\mathcal{E}}}

\newcommand{\mEm}{ \mathcal{E}_{\textrm{min}}}
\newcommand{\bphi}{{\bar{\phi}}}

\newcommand\pubdate{ }

\def\KICC{Kavli Institute for Cosmology and Institute of Astronomy,\\ Madingley Rd, Cambridge CB3 0HA, United Kingdom}

\def\Title#1{\begin{center} {\Large #1 } \end{center}}
\def\Author#1{\begin{center}{ \sc #1} \end{center}}
\def\Address#1{\begin{center}{ \it #1} \end{center}}

\newcommand\pubblock{\rightline{\begin{tabular}{l}\\
         \pubdate \end{tabular}}}

\title{K-Oscillons: Oscillons with Non-Canonical Kinetic Terms}
\author{Mustafa A. Amin}

\date{ }
\begin{document}

\begin{titlepage}
\pubblock

\vfill \Title{K-Oscillons: Oscillons with Non-Canonical Kinetic Terms}
\vfill \Author{Mustafa A. Amin\footnote{email: mamin@ast.cam.ac.uk}} \Address{\KICC}


\abstract{
Oscillons are long-lived, localized, oscillatory scalar field configurations. In this work we derive a condition for the existence of small-amplitude oscillons (and provide solutions) in scalar field theories with non-canonical kinetic terms. While oscillons have been studied extensively in the canonical case, this is the first example of oscillons in scalar field theories with non-canonical kinetic terms. In particular, we demonstrate the existence of oscillons supported solely by the non-canonical kinetic terms, without any need for nonlinear terms in the potential. In the small-amplitude limit, we provide an explicit condition for their stability in d+1 dimensions against long-wavelength perturbations. We show that for $d\ge 3$, there exists a long-wavelength instability which can lead to radial collapse of small-amplitude oscillons.}
\vfill
\end{titlepage}
\def\thefootnote{\fnsymbol{footnote}}
\setcounter{footnote}{0}

%




\section{Introduction}
Scalar fields with non-canonical kinetic terms are used ubiquitously in cosmology. They are especially prevalent in modeling of the inflaton (e.g. \cite{Silverstein:2003hf,Chen:2006nt}), dark energy  and modifications of gravity (e.g. \cite{Wagoner:1970vr, ArmendarizPicon:2000ah}). A significant amount of work has been done on their homogeneous evolution in an expanding universe and the evolution of linearized fluctuations about this homogeneous background. However, less attention has been paid to their spatially varying, nonlinear dynamics (e.g. \cite{Felder:2002sv,Barnaby:2004dz,Endlich:2010zj, Padilla:2010ir, Adam:2011gc}). In particular, localized, {\it time-dependent}, soliton-like solutions in theories with non-canonical kinetic terms have been rarely discussed. If such configurations exist, they would be novel objects from a mathematical standpoint. If they form a significant component of the energy fraction of the universe, they might have cosmological consequences.

In this paper we show that in a general class of scalar field theories with non-canonical kinetic terms and/or nonlinear potentials, there exist extremely long-lived, spatially localized, oscillatory field configurations called oscillons \cite{Bogolyubsky:1976yu, Gleiser:1993pt}\footnote{Oscillons are similar to Q-balls \cite{Coleman:1985ki} in that their existence has nothing to do with topology, however unlike Q-balls, oscillons do not have an exactly conserved charge.} . While oscillons in scalar field theories have been studied extensively in the literature (for e.g. \cite{Copeland:1995fq, McDonald:2001iv, Kasuya:2002zs,Gleiser:2004iy,Gleiser:2006te, Hindmarsh:2006ur, Saffin:2006yk, Farhi:2007wj, Hindmarsh:2007jb, Fodor:2009kf, Fodor:2009xw, Gleiser:2009ys,Amin:2010xe, Amin:2010jq,Hertzberg:2010yz,Gleiser:2010qt,Gleiser:2011xj,Amin:2011hj,Salmi:2012ta,Dutra:2012nj, Andersen:2012wg, Sfakianakis:2012bq, Gleiser:2012tu,Zhou:2013tsa}), every instance so far uses a nonlinear term in the potential and a canonical kinetic term. To the best of our knowledge, this is the first time their existence is being demonstrated in the presence of non-canonical kinetic terms. When the non-canonical {\it kinetic} terms are {\it significant}, we refer to the oscillons as  ``k-oscillons''\footnote{Scalar field dark energy with significant non-canonical kinetic terms is often referred to as {\it k-essense} (as opposed to quintessence).}. We derive the condition for their existence, including effects from the non-canonical kinetic term as well as the nonlinearity in the potential. Our results are general enough to include oscillons supported by nonlinear potentials, oscillons supported purely by the non-canonical kinetic terms as well as oscillons supported by a combination of both. The inclusion of non-canonical kinetic terms significantly expands the space of theories where oscillons can exist. 

Our analysis is done in a small-amplitude approximation, but is otherwise quite general. We consider  scalar field Lagrangians of the form
\Beq
\label{eq:lag1}
\mathcal{L}
&=T(X,\vp)-V(\vp),
\Eeq
in $d+1$ space-time dimensions where
\Beq
X&=-\frac{1}{2}\eta^{\alpha\beta}\partial_\alpha\vp\partial_\beta\vp.\\
\Eeq
$\eta_{\mu\nu}$ is a Minkowski space metric with the `mostly $+$' signature. The only restriction of $T$ and $X$ is that they can be written as
\Beq
\label{eq:lag2}
T(X,\vp)&=X+\xi_{2}X^2+\xi_3\vp X^2+\hdots\\
V(\vp)&=\frac{1}{2}\vp^2+\frac{\lambda_3}{3}\vp^3+\frac{\lambda_4}{4}\vp^4+\frac{\lambda_5}{5}\vp^5+\hdots\\
\Eeq
where all the field variables, space-time co-ordinates and coefficients have been made dimensionless using appropriate scalings. In the next section we motivate this Lagrangian and discuss scalings of the parameters and fields in terms of a mass and a cutoff scale. Anticipating a small-amplitude expansion, we have organized the series in terms of powers of the field and kept terms up to fourth order in the fields.  Note that terms of the form $\vp X$ and $\vp^2 X$ which should be included in the above expression can always be absorbed using a field re-definition. Furthermore, note that  the $X^2$ term cannot be eliminated using a field re-definition. Our choice of $T$ and $V$ also ensures that we recover a free, canonical scalar field theory when $X, \vp\rightarrow 0$. This form is general enough to cover a large class of scalar field theories of interest in cosmology including axions \cite{Weinberg:1978}, DBI inflation \cite{Silverstein:2003hf}, monodromy inflation \cite{Silverstein:2008sg,McAllister:2008hb}, k-essence \cite{ArmendarizPicon:2000ah} and scalar-tensor theories \cite{Wagoner:1970vr}.

For the Lagrangian discussed above, and in the small-amplitude, spherically symmetric case, we provide
\begin{itemize}
\item  a condition for the existence of oscillons in terms of a relationship between the first few coefficients in the series for $T$ and $V$,
\item explicit, controlled, analytic solutions in $1+1$ dimensions and approximate solutions in $3+1$ and higher dimensions,
\item a condition for their stability against long wavelength perturbations. The condition shows the presence of a long wavelength instability for $d\ge 3$ (in the small-amplitude approximation).
\end{itemize}
We also calculate the energy loss from these general oscillons due to an expanding background. 

For canonical kinetic terms, the authors of \cite{Gleiser:2010qt, Amin:2010xe, Amin:2010dc, Gleiser:2011xj, Amin:2011hj} show that oscillons are produced copiously at the end of inflation (as well as in phase transitions and bubble collisions, for e.g., \cite{Dymnikova:2000dy,Johnson:2011wt,Cheung:2012im}) and can dominate the energy density of the universe at that time (e.g. \cite{Amin:2011hj}). A similar situation is possible with a somewhat contrived model of dark energy as well \cite{Amin:2011hu}. Models with non-canonical kinetic terms and/or nonlinear potentials are well suited for amplifying fluctuations around a homogeneous oscillatory background field. This provides a natural mechanism to amplify quantum fluctuations existing at the end of inflation, possibly leading to the formation of large-amplitude oscillons. We will pursue this possibility in inflationary models with non-canonical kinetic terms in future work.  Indeed our original motivation for studying oscillons in non-canonical theories arose while trying to explore self-resonance  and preheating in the DBI scenario (see for e.g. \cite{Davis:2009wg, Bouatta:2010bp, Karouby:2011xs})

The rest of this paper is organized as follows. In the section 2 we motivate the Lagrangian discussed in the introduction, in section 3 we derive an oscillon solution in $d+1$ dimensions, in section 4 we discuss the stability of the oscillon solutions and in section 5 we summarize the main results and discuss directions for future work. 
\section{The effective Lagrangian}
In this section we discuss the motivation for the Lagrangian in the introduction. For the reader interested in oscillons from a mathematical standpoint, this section can be omitted without affecting the rest of the paper. From an effective field theory perspective, a general Lagrangian with non-canonical kinetic terms has the form 
\Beq
\mathcal{L}_\phi=a_1(\phi)X_\phi+a_2(\phi)\frac{X_\phi^2}{\Lambda^{d+1}}+\hdots -U(\phi),
\Eeq
where $X_\phi=-(1/2)\partial_{\bar{\mu}}\phi\partial^{\bar{\mu}}\phi$. We are assuming that only first derivatives of the field appear in the Lagrangian. Let $U(\phi)=(1/2)m^2\phi^2+\hdots$ and $\Lambda\gg m$ is the cutoff scale. Furthermore, we assume that a Taylor expansion exists for all $a_n(\phi)$ and $a_1(0)=1$. A field re-definition enacted via $d\bphi/d\phi=\sqrt{a_1(\phi)}$ yields 
\Beq
\mathcal{L}_{\bphi}=X_\bphi+b_2(\bphi)\frac{X_\bphi^2}{\Lambda^{d+1}}+\hdots -V(\bphi),
\Eeq
where $X_\bphi=-(1/2)\partial_{\bar{\mu}}\bphi\partial^{\bar{\mu}}\bphi$. We Taylor expand $\mathcal{L}$  around $(\bphi,X_\bphi)=(0,0)$ and assume that as $(X_\bphi,\bphi)\rightarrow (0,0)$ we recover a canonical free-field Lagrangian. This yields:
\Beq
\mathcal{L}=X_\bphi+\bar{\xi}_2\frac{X_\bphi^2}{\Lambda^{d+1}}+\hdots -\frac{1}{2}m^2\bphi^2-\frac{1}{3}m^{\frac{5-d}{2}}\bar{\lambda}_3\bphi^3-\frac{1}{4}m^{3-d}\bar{\lambda}_4\bphi^4+\hdots
\Eeq
Let us redefine the fields and space-time variables as follows
\Beq
\label{eq:scalings}
x^\mu&=m x^{\bar{\mu}},\\
\vp&=m^\frac{1-d}{2}\bar{\phi},\\
X&=m^{-(d+1)}X_{\bar{\phi}},\\
\xi_2&=\left(\frac{m}{\Lambda}\right)^{(d+1)}\bar{\xi}_2,\\
\lambda_n&=\bar{\lambda}_n.
\Eeq
With these redefinitions we get the Lagrangian we will use for the rest of the paper:
\Beq
\mathcal{L}&=T(X,\vp)-V(\vp)\\
&=\left[X+\xi_{2}X^2+\hdots \right] -\left[\frac{1}{2}\vp^2+\frac{\lambda_3}{3}\vp^3+\frac{\lambda_4}{4}\vp^4+\hdots\right],
\Eeq
where anticipating a small-amplitude expansion, we have only kept terms up to $4$th order in the field $\vp$. Note that the next order terms in the first and second brackets above would have the form $\xi_3\vp X^2$ and $(1/5)\lambda_5\vp^5$ respectively.  

Another way of arriving at the Lagrangian above is as follows (see for e.g. \cite{Tolley:2009fg}). Consider a Lagrangian with two fields in $3+1$ dimensions (the generalization to $d+1$ dimensions is straightforward):
\Beq
\mathcal{L}=-\frac{1}{2}\partial_{\bar{\mu}}\bphi\partial^{\bar{\mu}}\bphi-\frac{1}{2}\partial_{\bar{\mu}} \psi\partial^{\bar{\mu}} \psi-\frac{1}{2}\Lambda^2\psi^2-\frac{1}{2}m^2\bphi^2-\frac{1}{3}m \bar{\lambda}_3\bphi^3-\frac{1}{4}\bar{\lambda}_4\bphi^4-\sqrt{\frac{\bar{\xi}_2}{2}}\frac{\psi}{\Lambda}\partial_{\bar{\mu}}\bphi\partial^{\bar{\mu}}\bphi+\hdots
\Eeq
with $\Lambda\gg m$. The heavy field will sit at the minimum of its effective potential, with the corresponding field value given by
\Beq
\psi_*=-\sqrt{\frac{\bar{\xi}_2}{2}}\frac{\partial_{\bar{\mu}}\bphi\partial^{\bar{\mu}}\bphi}{\Lambda^3}.
\Eeq
Substituting into the original Lagrangian and setting the kinetic term of the heavy field to zero, we have
\Beq
\mathcal{L}&=-\frac{1}{2}\partial_{\bar{\mu}}\bphi\partial^{\bar{\mu}}\bphi+\bar{\xi}_2\frac{\left(\partial_{\bar{\mu}}\bphi\partial^{\bar{\mu}}\bphi\right)^2}{4\Lambda^4}-\frac{1}{2}m^2\bphi^2-\frac{1}{3}m \bar{\lambda}_3\bphi^3-\frac{1}{4}\bar{\lambda}_4\bphi^4+\hdots\\
&=X_\bphi+\bar{\xi}_2\frac{X_\bphi^2}{\Lambda^4}-\frac{1}{2}m^2\bphi^2-\frac{m}{3}\bar{\lambda}_3\bphi^3-\frac{1}{4}\bar{\lambda}_4\bphi^4+\hdots
\Eeq
With appropriate scalings of space-time and fields with mass $m$ and $\Lambda$ defined in equation \eqref{eq:scalings}, we recover the effective Lagrangian discussed above and in the introduction (see equation \eqref{eq:lag2}).

Note that if $\bar{\xi}_2$ and $\bar{\lambda}_n$ are order one, we expect $\xi_2\ll \lambda_n$ and $\xi_2\ll1$ if $\Lambda\gg m$. However, this need not always be the case. For example, in the DBI case $\mathcal{L}=f^{-1}(\vp)\left[1-\sqrt{1-2f(\vp)X}\right]-V(\vp)$\cite{Silverstein:2003hf} yields $\xi_2=1/2$. For the rest of the paper we do not make any particular assumptions about the sizes of $\xi_2$ and $\lambda_n$, apart from assuming that they are not much larger than unity. 


\section{Oscillons in $d+1$ dimensions}
Let us begin with the equations of motion associated with the Lagrangian presented in the introduction and discussed in the previous section:
 \Beq
 \label{eq:EOM}
\Box\vp+\frac{\partial_X^2 T}{\partial_X T}\partial_\mu\vp\partial^\mu X+\frac{\partial_X\partial_\vp T}{\partial_X T}\partial_\mu\vp\partial^\mu \vp=\frac{\partial_\vp V-\partial_\vp T}{\partial_X T}.
\Eeq
We are  interested in small-amplitude, radially symmetric, spatially localized, oscillatory (in time) solutions. First, we rescale the time and space variables by a small parameter $\epsilon$ as follows:
\Beq
\tau&=\sqrt{1-\epsilon^2}t,\\
\rho&=\epsilon r.
\Eeq
and expand the solution as a series in $\epsilon$ (a possibly asymptotic one):
\Beq
\label{eq:phie}
\vp(t,r)=\epsilon \phi_1(\tau,\rho)+\epsilon^2\phi_2(\tau,\rho)+\epsilon^3\phi_3(\tau,\rho)+\hdots
\Eeq
Plugging the above scalings and form of the solution into the equation of motion \eqref{eq:EOM}, and collecting terms to lowest order in $\epsilon$ we get
\Beq
&\partial_{\tau\tau}\phi_1+\phi_1=0,\\
\Longrightarrow\quad&\phi_1(\tau,\rho)=f(\rho)\cos \tau.
\Eeq
where we assumed that $\partial_\tau\phi_1(0,\rho)=0$. At second order in $\epsilon$ we get
\Beq
&\partial_{\tau\tau}\phi_2+\phi_2=-\frac{1}{2}\lambda_3f^2(\rho)\left[1+\cos 2\tau\right],\\
\Longrightarrow\quad&\phi_2(\tau,\rho)=\frac{1}{6}\lambda_3f^2(\rho)\left[-3+2\cos\tau+\cos 2\tau\right].
\Eeq
where we assumed that $\phi_2(0,\rho)=\partial_\tau\phi_2(0,\rho)=0$. 
At the next order in $\epsilon$, we get
\Beq
&\partial_{\tau\tau}\phi_3+\phi_3\\
&=\left[\partial_\rho^2 f+\frac{(d-1)}{\rho}\partial_\rho f-f+\frac{3}{4}\left(\xi_2-\lambda_4+\frac{10}{9}\lambda_3^2\right)f^3\right]\cos\tau\\
&\quad+\left[\hdots\right]\cos 2\tau\\
&\quad+\left[\hdots\right]\cos3\tau.
\Eeq
If the coefficient of the $\cos \tau$ term is non-zero, $\phi_3$ would grow linearly with time, inconsistent with the time-periodic solution we are looking for. \footnote{Note that the $[\hdots]\cos 2\tau$ and $[\hdots]\cos 3\tau$ terms will yield a periodic solution for $\phi_3$.  This solution can then be used in calculating terms at the next order, just as the periodic solution for $\phi_2$ played a role in the $\phi_3$ equation. This pattern extends to all orders.} To avoid linear resonance, we need to set the coefficient of $\cos \tau$ to zero. This in turn yields the equation for the profile $f(\rho)$:
\Beq
\label{eq:profile}
\partial_\rho^2 f(\rho)+\frac{(d-1)}{\rho}\partial_\rho f(\rho)-f(\rho)+\frac{3}{4}\Delta f^3(\rho)=0.
\Eeq
where for future convenience we have defined 
\Beq
\Delta\equiv \xi_2-\lambda_4+\frac{10}{9}\lambda_3^2,
\Eeq
whose sign will turn out to determine whether oscillons exist or not. 

This profile equation {\eqref{eq:profile}} is valid in any dimension $d$, but can be solved exactly for $d=1$. Let us consider the $d=1$ case first. 
\subsection{1+1 dimensional oscillons}
For $d=1$ the profile equation becomes
\Beq
\partial_\rho^2 f(\rho)-f(\rho)+\frac{3}{4}\Delta f^3(\rho)=0.
\Eeq
One can think of the above equation as describing the motion of a particle in a potential
\Beq
\mathcal{U}(f)&=-\frac{f^2}{2}+\frac{3}{16}\Delta f^4.
\Eeq
The energy associated with this motion is conserved, and given by
\Beq
\mathcal{E}=\mathcal{U}(f_0)=\frac{(\partial_\rho f)^2}{2}+\mathcal{U}(f),
\Eeq
where we have used $\partial_\rho f(0)=0$ and $f(0)\equiv f_0$. Now since the solutions are localized, the energy $\mathcal{E}=\mathcal{U}(f_0)=0$. This immediately, yields
\Beq
f_0=\sqrt{\frac{8}{3\Delta}}.
\Eeq
For a localized solution to exist, we need
\Beq
\boxed{\,\,\Delta=\xi_2-\lambda_4+\frac{10}{9}\lambda_3^2>0.\,\,}
\Eeq
This is one of our main results.  Before moving on to solving the profile equation, we pause to discuss $\Delta$ in a bit more detail. If the non-canonical terms are absent ($\xi_2=0$), we need the usual ``opening up of the potential" condition: $-\lambda_4+(10/9)\lambda_3^2>0$ to get oscillons. More importantly, note that for a quadratic potential (i.e. $\lambda_n=0$), the non-canonical terms are sufficient to yield oscillons. For example if $\lambda_n=0$, then $\xi_2>0$ is sufficient. It is also worth noting that in models with non-canonical kinetic terms, the sound speed differs from $1$. For the model under consideration, the sound speed is 
\Beq
c_s^2
&=\left(1+2X\frac{\partial_X^2 T}{\partial_X T}\right)^{-1}\\
&=1-4\xi_2 X+\hdots
\Eeq
Thus, for $\lambda_n=0$, the condition for having k-oscillons is the same as the sound speed being less than 1. 

Now, let us get back to solving the equation for the profile $f(\rho)$. Using $f(0)=f_0$ derived above and integrating $(\partial_\rho f)^2/{2}+\mathcal{U}(f)=0$ yields 
\Beq
f(\rho)=\sqrt{\frac{8}{3\Delta}}\textrm{sech}( \rho).
\Eeq
Explicitly in terms of the original variables:
\Beq
\boxed{\vp(t,r)=\epsilon\sqrt{\frac{8}{3\Delta}}\textrm{sech}\left(\epsilon r\right)\cos\left(\sqrt{1-\epsilon^2}t\right)+\mathcal{O}[\epsilon^2]}
\Eeq
where $\Delta=\xi_2-\lambda_4+\frac{10}{9}\lambda_3^2$.
This is our second main result: a small-amplitude oscillon in $1+1$ dimensions in theories with non-canonical kinetic terms. This solution has the same functional form as the canonical small-amplitude oscillon, apart from the appearance of $\xi_2$ in $\Delta$. One can go beyond the leading order as well:
\Beq
\label{eq:sol1D}
\vp(t,r)
&=\epsilon\sqrt{\frac{8}{3\Delta}}\textrm{sech}\rho \cos\tau+\epsilon^2\frac{4\lambda_3}{9\Delta}\textrm{sech}^2\rho\left[-3+2\cos \tau+\cos 2\tau\right]\\
\quad&+\mathcal{O}[\epsilon^3]
\Eeq
where $\rho=\epsilon r$ and $\tau=\sqrt{1-\epsilon^2}t$. Note that at second order, only the odd term contributes. If the Lagrangian had $\vp\rightarrow-\vp$ symmetry, the the correction is higher order. 

\subsection{$3+1$ dimensional oscillons}
For $d=3$, the profile equation becomes
\Beq
\partial_\rho^2 f(\rho)+\frac{2}{\rho}\partial_\rho f(\rho)-f(\rho)+\frac{3}{4}\Delta f^3(\rho)=0.
\Eeq
We can view the above equation as describing the motion of particle in the presence of a potential $\mathcal{U}(f)$ as in the $1+1$ D case, but now we have a `friction term' of the form $(2/\rho)\partial_\rho f(\rho)$. As a result, the energy 
\Beq
\mathcal{E}(\rho)=\frac{(\partial_\rho f)^2}{2}+\mathcal{U}(f)
\Eeq
is no longer conserved. It changes with $\rho$ as
\Beq
\partial_\rho\mathcal{E}(\rho)=-\frac{2}{\rho}(\partial_\rho f)^2.
\Eeq
With the requirement that the solution is ÒlocalizedÓ, we need $ \mathcal{E}\rightarrow 0$ as $\rho\rightarrow \infty$. Requiring that the solution is smooth at $\rho=0$ implies $\partial_\rho f(0) = 0$. This implies that for a localized solution we must have $\mathcal{E}(0)=\mathcal{U}(f_0)\ge 0$. Numerically, one finds a localized solution for \footnote{more precisely  \Beq
f_0\approx 3.06699\times \sqrt{\frac{8}{3\Delta}}
\Eeq}
\Beq
f_0\approx \sqrt{\frac{24}{\Delta}}.
\Eeq
Thus the solution will have the form:
\Beq
\vp(t,r)\approx \epsilon f_0F(\epsilon r) \cos\left(\sqrt{1-\epsilon^2}t\right)+\mathcal{O}[\epsilon^2]
\Eeq
The profile $F(\rho)$ looks `sech' like with $F(0)=1$. An excellent approximation (at the few $\%$ level in the `core' region, deteriorating in the tails) is given by
\Beq
F(\rho)=\sqrt{\textrm{sech}\,\left(f_0^{\frac{2}{3}(2\Delta)^{1/3}} \rho\right)}\quad\quad 1\lesssim \Delta \lesssim \textrm{few} 
\Eeq
An identical analysis can be carried out in $d\ne 1,3$ and we do not repeat the derivation here.


\section{Stability}

We now turn to the question of stability of the oscillons against perturbations. To this end, we linearize the equation of motion \eqref{eq:EOM} around the small-amplitude oscillons solutions $\vp_{osc}$ as follows:
\Beq
\delta\ddot{\vp}-6\xi_2\vp_{osc}\dot{\vp}_{osc}\delta\dot{\vp}+\left[-\left(1-2\xi_2\dot{\vp}_{osc}^2\right)\nabla^2+1+2\lambda_3\vp_{osc}+3\lambda_4\vp_{osc}^2-3\xi_2\dot{\vp}_{osc}^2\right]\delta\vp=0.
\Eeq
We have used the following to arrive at the above equation:
\Beq
&\mathcal{O}[\vp_{osc}]\sim\mathcal{O}[\dot{\vp}_{osc}]\sim\epsilon,\\
&\mathcal{O}[|\nabla{\vp}_{osc}|]\sim\epsilon^2,
\Eeq
and kept terms up to order $\epsilon^2$. The $\xi_2\vp_{osc}^2\nabla^2\delta\vp$ term has to be kept since we have not (yet) restricted ourselves to long wavelength perturbations. Let us remove the term with the linear derivative by redefining the $\delta\vp$ as follows:
\Beq
\delta\vp=\chi \exp\left[3\xi_2\int_0^t ds{\dot{\vp}_{osc}\vp_{osc}}\right].
\Eeq
With this redefinition, the equation of motion becomes
\Beq
\label{eq:Pert0}
\ddot{\chi}+\left[-\left(1-2\xi_2\dot{\vp}_{osc}^2\right)\nabla^2+1+2\lambda_3\vp_{osc}-3(\xi_2-\lambda_4){\vp}_{osc}^2\right]\chi=0.
\Eeq

The solutions $\vp_{osc}$ are periodic in time. Hence stability can be determined via a Floquet analysis. It is tempting to Fourier transform the above equation and try to determine the Floquet exponents (growth-rate) mode by mode. However, since the background varies in space, the Fourier modes do not decouple. A stability analysis in Fourier space, while possible (see \cite{Hertzberg:2010yz}), is numerically intensive, especially in higher dimensions. We {\it do not} follow this approach here. Instead we will carry out a stability analysis in position space, but restrict ourselves to perturbations which vary on length scales comparable to the size of the oscillon. Even without considering short wavelengths, we will show that there exists an important instability in dimensions $\ge 3$. 
\subsection{Long-wavelength stability analysis}
When considering perturbations with wavelengths comparable to the size of the oscillon, we can drop the $\xi_2\dot{\vp}_{osc}^2\nabla^2\delta\vp\sim \epsilon^4$ term in equation \eqref{eq:Pert0} since it is higher order in $\epsilon$. This yields
\Beq
\label{eq:Pert}
\ddot{\chi}+\left[-\nabla^2+1+2\lambda_3\vp_{osc}-3(\xi_2-\lambda_4){\vp}_{osc}^2\right]\chi=0.
\Eeq
At this point, the equation of motion for linearized fluctuations is identical to that in the canonical case, apart from the coefficient $\xi_2$. From now onwards, our long wavelength stability calculation closely follows the one by Amin and Shirokoff \cite{Amin:2010jq}, where we related oscillon stability to the stability criterion derived by Vakhitov and Kolokolov \cite{Vakhitov:1975} in the context of light focusing in a nonlinear medium. Here, apart from showing this relationship in the context of non-canonical oscillons, we also provide a pedagogical proof of the stability criterion itself in an Appendix (not provided in \cite{Amin:2010jq}).

Recall that the oscillon solutions have the form
\Beq
\vp_{osc}(t,r)&=\epsilon f(\rho)\cos(\tau)\\
&\quad+\epsilon^2\frac{\lambda_3}{6}f^2(\rho)\left[-3+2\cos\tau+\cos 2\tau\right]+\hdots
\Eeq
where $\rho=\epsilon r$, $\tau=\sqrt{1-\epsilon^2}t$ and $f$ is the radial oscillon profile satisfying equation \eqref{eq:profile}. 

We are interested in determining the stability of the above oscillons to perturbations with wavelengths comparable to the size of the oscillons. With this in mind, let us define a scaled spatial co-ordinate $\trho=\te r$ where $\te=\epsilon/\sqrt{\alpha}$ with $\alpha$ being an order 1 parameter. We expect the most unstable perturbations to oscillate at the oscillon frequency: $\sqrt{1-\epsilon^2}$ with a `slowly-varying', time-dependent envelope driven by the oscillating background. To capture the time dependence of the envelope we define a slow time $T=\te^2t$ which is to be treated independently from $\tau$. Note that although $\mathcal{O}[\te]\sim\mathcal{O}[\epsilon]$, they are independent. $\epsilon$ plays the role of determining the oscillon solution whereas $\te$ is used for analyzing the stability about this solution. This distinction is made explicit via the introduction of the $\alpha$ parameter. With these definitions we have
\Beq
\frac{d^2}{dt^2}&=(1-\epsilon^2)\partial_\tau^2+\te^2\sqrt{1-\epsilon^2}2\partial_T\partial_\tau+\te^4\partial_T^2\\
&\approx\partial_\tau^2+\epsilon^2\left(\frac{2}{\alpha}\partial_T\partial_\tau-\partial_\tau^2\right),\\
\nabla^2&=\te^2\tilde{\nabla}^2=\frac{\epsilon^2}{\alpha}\tilde{\nabla}^2.
\Eeq
Furthermore, let us expand the perturbation $\chi$ in powers of $\epsilon$ as follows:
\Beq
\chi
&=\chi_0+\te\chi_1+\te^2\chi_2+\hdots\\
&=\chi_0+\frac{1}{\sqrt{\alpha}}\epsilon\chi_1+\frac{1}{\alpha}\epsilon^2\chi_2+\hdots\\
\Eeq

We now substitute the space-time scalings, the $\epsilon$ expansion of the perturbation and the oscillon solution into the EOM of the perturbation \eqref{eq:Pert}. Collecting terms order by order in $\epsilon$ we have (up to order $\epsilon^2$)
\Beq
\label{eq:PertOrder}
&\left[\partial_\tau^2+1\right]\chi_0=0,\\
&\left[\partial_\tau^2+1\right]\chi_1+\left[2\lambda_3\Phi\cos\tau\right]\chi_0=0,\\
&\left[\partial_\tau^2+1\right]\chi_2+\left[2\lambda_3\Phi\cos\tau\right]\chi_1
+\left[2\partial_\tau\partial_T-\tilde{\nabla}^2-\alpha\partial_\tau^2-\frac{3}{2}\left(\xi_2-\lambda_4+\frac{2}{3}\lambda_3^2\right)\Phi^2\right.\\
&\quad\quad\quad\quad\quad\quad\quad\quad\quad\quad\quad\quad\quad\quad+\left.\frac{2}{3}\lambda_3^2\Phi^2\cos\tau-\frac{3}{2}\left(\xi_2-\lambda_4-\frac{2}{9}\lambda_3^2\right)\Phi^2\cos2\tau\right]\chi_0=0,
\Eeq
where we have defined
\Beq
\Phi(\alpha,\trho)\equiv \sqrt{\alpha}f(\sqrt{\alpha}\trho)=\sqrt{\alpha}f(\rho).
\Eeq
Note that treating $\mathcal{O}[\tilde{\nabla}^2\chi_0]\sim \mathcal{O}[\chi_0]$ we are restricting ourselves to perturbations that vary spatially on the scale of $\epsilon$. The first equation of \eqref{eq:PertOrder} yields
\Beq
\chi_0=u(T,\trho)\cos\tau+v(T,\trho)\sin\tau.\\
\Eeq
Substituting $\chi_0$ into the second equation and solving for $\chi_1$ we get
\Beq
\chi_1=\frac{1}{3}\lambda_3\Phi\left[u(-3+\cos 2\tau)+v \sin2\tau\right].
\Eeq
where  we have ignored the homogeneous solution of $\chi_1$. Finally, substituting $\chi_0$ and $\chi_1$ into the third equation we get
\Beq
\left[\partial_\tau^2+1\right]\chi_2=&-\left[2\partial_Tv-\left(\tilde{\nabla}^2-\alpha+\frac{9}{4}\Delta\Phi^2\right)u\right]\cos\tau\\
&-\left[-2\partial_Tu-\left(\tilde{\nabla}^2-\alpha+\frac{3}{4}\Delta\Phi^2\right)v\right]\sin\tau\\
&+\left[\hdots\right]\cos3\tau+\left[\hdots\right]\sin3\tau,
\Eeq
where $\Delta=\xi_2-\lambda_4+(10/9)\lambda_3^2$ (the combination which appears in the oscillon solution). Avoiding secular growth requires
\Beq
\label{eq:system}
\partial_Tu&=H_1 v,\\
\partial_Tv&=-H_2u,\\
\Eeq
where
\Beq
H_1&\equiv-\frac{1}{2}\left(\tilde{\nabla}^2-\alpha+\frac{3}{4}\Delta\Phi^2\right),\\
H_2&\equiv-\frac{1}{2}\left(\tilde{\nabla}^2-\alpha+\frac{9}{4}\Delta\Phi^2\right).\\
\Eeq
We are interested in the eigenvalues of the above linear system. Let $v(T,\trho)=e^{\Omega T}v_e(\trho)$ and $u(y,T)=e^{\Omega T}u_e(\trho)$ where $(u_e,v_e)$ is an eigenvector of equation \eqref{eq:system}. Substituting $(u,v)=e^{\Omega T}(u_e,v_e)$ into our linear system, we get
\Beq
\Omega u_e&=H_1v_e,\\
\Omega v_e&=-H_2 u_e.
\Eeq
Equivalently
\Beq
\label{eq:eigen}
-\Omega^2 u_e&=H_1H_2u_e,\\
-\Omega^2 v_e&=H_2H_1v_e.
\Eeq
Since $H_1$ and $H_2$ are real operators, the eigenvalues $-\Omega^2$ are real, that is $\Omega$ is purely real or imaginary. There exist exponentially growing modes if and only if $\min[-\Omega^2]<0$. Hence, we now try to determine $\min[-\Omega^2]$.

In the above form, our problem becomes similar to that of light focusing in a nonlinear medium as analyzed by Vakhitov and Kolokolov \cite{Vakhitov:1975}. Following their techniques, we will show in the Appendix that
\Beq
&\textrm{sign}\left[\min[-\Omega^2]\right]=\textrm{sign}\left[\frac{d N}{d\alpha}\right],\\
\Eeq
where
\Beq
N&\equiv\int \Phi^2(\alpha,\trho) d^d \trho\\
&= \alpha\int f^2(\sqrt{\alpha}\trho)d^d \trho\\
&=\alpha^{(1-d/2)}\int f^2(\rho)d^d\rho\\
&\propto \alpha^{(1-d/2)}.\\
\Eeq
The proof of the relationship between the sign of $\frac{d N}{d\alpha}$ and $\textrm{sign}\left[\min(-\Omega^2)\right]$ that we used in the first line above is somewhat involved, which is why we have moved it to an Appendix. Here, we discuss the important and interesting consequences of the result.

For $d>2$, we will have $dN/d\alpha<0$ and thus $\textrm{sign}\left[\min[-\Omega^2]\right]<0$. Whereas for $d\le 2$ we have $dN/d\alpha\ge 0$ and thus $\textrm{sign}\left[\min[-\Omega^2]\right]>0$. Evidently, our oscillons are stable against long wavelength perturbations in $d=1,2$ but not so in $d>2$. This is confirmed by our numerical simulations. For $\xi_2=\lambda_3=0$, and in the small-amplitude limit, this reduces to the result in \cite{Amin:2010jq}. 

Recall that $\alpha$ is merely a scaling of $\epsilon$ in the oscillon solution. In terms of $\epsilon$, the stability condition is as follows: Oscillons are stable if and only if
\Beq
\label{eq:StabilityCond}
\boxed{\frac{dN}{d\epsilon}>0,}
\Eeq
where
\Beq
N\equiv \epsilon^2\int f^2(\epsilon r)d^d r=\epsilon^{2-d}\int f^2(\rho)d^d \rho\propto \epsilon^{d-2},
\Eeq
where $f$ is the oscillon profile, that is $\vp_{osc}=\epsilon f(\epsilon r)\cos \sqrt{1-\epsilon^2}t$. We have numerically verified the existence of a long wavelength instability for $d\ge 3$. 

We stress that the stability criterion \eqref{eq:StabilityCond} is applicable in the small-amplitude limit. More precisely it is applicable when a single frequency solution is a good approximation to the true solution. We now connect the above result to some related work in the literature. 

In the example of \cite{Amin:2010jq} with canonical kinetic terms, unlike our discussion here, the coefficient of the  $\vp^6$ term was assumed to be unusually large. This allowed for an (approximate) single frequency solution for the entire allowed amplitude range, which in turn allowed for the derivation of the same stability criterion derived above. However, unlike the above case, in the large $\vp^6$ case, $N$ was a non-monotonic function of $\epsilon$ in $3+1$ dimensions, allowing stable solutions to exist at large amplitudes. We also note that a similar stability criterion in term of the oscillon energy was also conjectured in \cite{Fodor:2009xw} based on numerical results in a massless dilaton + scalar field oscillon. While no general stability condition exists for the general, large-amplitude case, stability and lifetime of large amplitude oscillons is often investigated using Gaussian initial profiles with varying widths and amplitudes. For a flavor of such investigations see for example \cite{Gleiser:2009ys,Salmi:2012ta}. 

\subsection{Radiating Tails in an Expanding Universe}
While we have discussed the stability of our oscillons against `external' long-wavelength perturbations, even without external perturbations, oscillons are not exactly stable. Similar to the canonical case, our more general oscillons possess a radiating tail which we expect to be highly suppressed \cite{Segur:1987mg,Fodor:2009kf}, with a decay rate $\sim e^{-1/\epsilon}$\footnote{There exist scenarios where there are no radiating tails \cite{Arodz:2011zm}. We thank an anonymous referee for pointing this out.}. Nevertheless, in an expanding universe, this tail can be significantly enhanced \cite{Graham:2006xs, Farhi:2007wj, Amin:2010jq}.\footnote{A quantum treatment of the radiation will also increase the decay rate, with the decay rate becoming a power law in $\epsilon$ \cite{Hertzberg:2010yz}.}
We briefly sketch out the energy loss due to this radiating tail in an expanding universe below. 

For simplicity we will only consider the case in $1+1$ dimensions. In local co-ordinates, the metric for a homogeneous and isotropic expanding space can be written as (space and time are measured in units of $m^{-1}$)
\Beq
ds^2=-(1-x^2H^2)dt^2+(1-x^2H^2)^{-1}dx^2,
\Eeq
where we assume that $H=$ constant and $\mathcal{O}[H/m]= \mathcal{O}[\epsilon^2]$. In this case the solution takes the following form (following the technique used in \cite{Farhi:2007wj}):
\Beq
\vp(x,t)\approx\epsilon\sqrt{\frac{8}{3\Delta}}\textrm{sech}\left(\epsilon x\right)\cos\left(\sqrt{1-\epsilon^2}t\right)\quad\quad x\ll \frac{\epsilon}{H},
\Eeq
and
\Beq
\vp(x,t)\approx\epsilon^{3/2}e^{-\frac{\pi\epsilon^2}{2H}}2\sqrt{\frac{8}{3Hx\Delta}}\cos\left(\sqrt{1-\epsilon^2}t-\frac{1}{2}x^2H\right)\quad\quad \frac{\epsilon}{H}\ll x\ll \frac{1}{H}.
\Eeq
which leads to an energy loss (averaged over time) given by 
\Beq
\frac{dE_{osc}}{dt}\approx \epsilon^3\frac{32}{3\Delta}e^{-\frac{\pi\epsilon^2}{H}}\quad\quad\quad\quad\frac{\epsilon}{H}\ll x\ll \frac{1}{H}.
\Eeq
 Our analysis appropriately generalizes the result of \cite{Farhi:2007wj}. The $\Delta$ appearing above contains $\xi_2$ from the non-canonical kinetic term along with $\lambda_3$ and $\lambda_4$ whereas in  \cite{Farhi:2007wj}, $\Delta=-\lambda_4$. Note that this analysis in only valid when $m\gg H$. The energy loss while enhanced compared to the Minkowski case, can still lead to lifetimes $\gg H^{-1}$.



\section{Discussion}
In this paper we have shown that oscillons can exist in a significantly larger class of scalar field theories than previously shown. For a rather general class of scalar field Lagrangians of the form \eqref{eq:lag1} and \eqref{eq:lag2}, we have demonstrated  the following:
\begin{itemize}
\item For small-amplitude oscillons to exist, $\Delta=\xi_2-\lambda_4+(10/9)\lambda_3^2>0$ where $\xi_2$ is the coefficient of the non-canonical part of the kinetic term. 
\item The oscillon solutions in $d+1$ dimensions have the form $\vp(t,r)=\epsilon f(\epsilon r)\cos\sqrt{1-\epsilon^2} t+\mathcal{O}[\epsilon^2]$ where $f(\epsilon r)$ is the radial profile. In $1+1$ dimensions, $f(\epsilon r)=\sqrt{8/(3\Delta)}\textrm{sech}(\epsilon r)$. We also provided an approximate form for $f(\epsilon r)$ in $3+1$ dimensions. These solutions are identical to those in theories with canonical kinetic terms, apart from the appearance of $\xi_2$ in $\Delta$.
\item The solutions are stable against long wavelength perturbations if and only if $dN/d\epsilon>0$, where $N=\epsilon^{2-d}\int f^2(\rho)d^d\rho$. 
\end{itemize}
We have also calculated the energy loss from oscillons due to an expanding background. 

There are a number of natural extensions of our results. The stability criterion above is related to long-wavelength instabilities, which we believe to be the most dangerous instabilities. However, as discussed in the stability section, a calculation of the `Floquet' instability rates at shorter wavelengths while numerically intensive, is also possible. A further detailed investigation of the suppressed radiating tail, effects of expansion on lifetimes,  `Floquet instabilities' as well as a quantum treatment for these non-canonical oscillons would be interesting. 

We have concentrated on the small-amplitude regime in this paper. However, large-amplitude oscillons (with large energies) in $3+1$ dimensions are interesting due to their possible relevance in cosmology (see for e.g \cite{Gleiser:2010qt, Amin:2010xe, Amin:2010dc, Gleiser:2011xj, Amin:2011hj}). In addition, in $3+1$ dimensions, single field, small-amplitude oscillons can collapse due to perturbations with wavelengths comparable to the size of the oscillons. As we move to larger amplitudes this instability can disappear (e.g. \cite{Amin:2010jq}) . Thus an analysis of the large-amplitude case is certainly worth pursuing. However, moving to large amplitudes also requires a larger number of terms in $T$ and $V$, which in turn requires a case by case analysis of the solutions and their stability. It is of course possible to analyze them numerically. Although we have not presented the results here, we have analysed large-amplitude oscillons in the DBI Lagrangian: $\mathcal{L}=f^{-1}(\vp)\left[1-\sqrt{1-2f(\vp)X}\right]-\vp^2/2$, with rather intriguing dynamics appearing at large amplitudes \cite{Pearce:2013}. To keep our analysis as general as possible, and in an effort to present analytic rather than numerical results, we have restricted ourselves to the small-amplitude case in this paper. An analysis of large-amplitude k-oscillons and their implications in a cosmological context is in progress \cite{Pearce:2013}. 

\section{Acknowledgements}
We thank Michael Pearce for numerical simulations of large amplitude DBI oscillons, Mark Hertzberg for help with the small-amplitude expansion in the DBI case as well as a number of useful suggestions, David Shirokoff for help in understanding the proof of Vakhitov and Kolokolov,  and Neil Barnaby and Navin Sivanandam for comments on the effective Lagrangian. We thank Alan Guth, David Kaiser, Ruben Rosales and members of the Density Perturbation Group at MIT (2012) for helpful discussions during the early stages of this project. We also thank Alex Hall for a careful reading of the manuscript, Richard Easther, Ed Copeland, Marcelo Gleiser and Shuang-Yong Zhou for comments, and importantly, Eugene Lim for suggesting the name {\it k-oscillons}. We acknowledge support from a Kavli Fellowship.



\section*{Appendix}
\subsection*{Proof of $\textrm{sign}[\min[-\Omega^2]=\textrm{sign}[dN/d\alpha]$}
Our proof will closely follow the stability analysis of Vakhitov and Kolokolov presented in the context of light focusing in a nonlinear medium \cite{Vakhitov:1975}.  We will need two technical results regarding the Hermitian operators $H_1$ and $H_2$: 
\begin{enumerate}[(a)]
\item $\langle u_e|H_1^{-1}|u_e\rangle$ is positive definite for $\Omega\ne0.$ \footnote{where $\langle\hdots\rangle=\int\hdots d^d\trho$ and we are using the usual bra-ket notation}
\item $H_2$ has only one bounded eigenmode with a negative eigenvalue, all other eigenvalues for radially symmetric eigenmides are greater than zero, and the lowest angular eigenmode has zero eigenvalue. 
\end{enumerate}
We will assume these to be true for the moment and proceed with the proof. After the proof, we justify (a), but for (b) we refer the reader to \cite{Vakhitov:1975}.

We believe that some of the statements in the rest of the proof can be understood more readily based on our experience and intuition with single particle quantum mechanics. In particular, $H_1$ (and $H_2$) can be thought of as a non-relativistic Hamiltonian of a particle in a finite, radially symmetric potential in $d$ spatial dimensions. Hence we use language from quantum mechanics where appropriate. The energies and eigenstates of $H_1$ are denoted by $\{E_{\beta},\Psi_\beta\}$ and those of $H_2$ by $\{\mE_{\gamma},\psi_\gamma\}$

From equation \eqref{eq:eigen} we get $-\Omega^2\langle\Phi|u_e\rangle=\langle H_1\Phi |H_2u_e\rangle=0$. The first equality uses the Hermitian nature of $H_1$ whereas the second follows because $H_1\Phi=0$ is simply the oscillon profile equation \eqref{eq:profile}. Hence for $\Omega\ne 0$, $\langle\Phi|u_e\rangle=0$. If $\langle u_e|H_1^{-1}|u_e\rangle$ is nonzero, we can rewrite the first equation in \eqref{eq:eigen} as
\Beq
-\Omega^2=\frac{\langle u_e|H_2|u_e\rangle}{\langle u_e|H_1^{-1}|u_e\rangle}\quad\quad\textrm{with}\quad\quad\langle\Phi|u_e\rangle=0.\\
\Eeq
Now, since $\langle u_e|H_1^{-1}|u_e\rangle$ is positive definite based on (a) stated above, we have
\Beq
\label{eq:sign}
\textrm{sign}[\min[-\Omega^2]]&=\textrm{sign}[\min\left[{\langle u_e|H_2|u_e\rangle}\right]],\\
\Eeq
with 
\Beq
\langle\Phi|u_e\rangle&=0\quad\textrm{and}\quad\langle u_e|u_e\rangle&=1.\\
\Eeq
We introduce Lagrange multipliers $\mE$ and $\beta$ to minimize ${\langle u_e|H_2|u_e\rangle}$ subject to the above constraints:
\Beq
\fF[u_e,\mE,\beta]=\langle u_e|H_2|u_e\rangle+\mE\left(\langle u_e|u_e\rangle-1\right)+\beta \langle\Phi|u_e\rangle.
\Eeq
The extremum of $\fF$ is obtained if $u_e$ satisfies
\Beq
\label{eq:umin}
&H_2u_e=\mE u_e+\beta \Phi.
\Eeq
Moreover, the minimum value of ${\langle u_e|H_2|u_e\rangle}$ is given by the smallest eigenvalue of $H_2$ consistent with $\langle\Phi|u_e\rangle=0$. Let $\mEm$ denote this minimum eigenvalue. Then
\Beq
\textrm{sign}[\min[-\Omega^2]]=\textrm{sign}[\min\left[{\langle u_e|H_2|u_e\rangle}\right]]=\textrm{sign}[\mEm].
\Eeq
We will now try to determine the sign of $\mEm$. As mentioned earlier, one can also think of $H_2$ as the Hamiltonian of a non-relativistic particle in a finite potential well $V_2(\trho)=(\alpha/2)\left[1-(9/4)\textrm{sech}^2(\sqrt{\alpha}\trho)\right]$. Let $\{\psi_\gamma\}$ be the eigenstates of $H_2$ with energies $\mE_\gamma$. Let us expand $u_e$ and $\Phi$ in terms of these eigenstates as $\Phi=\sum_{\gamma=0} a_\gamma \psi_\gamma$ and $u_e=\sum_{\gamma=0} b_\gamma \psi_\gamma$. Plugging these into \eqref{eq:umin} and using  $\langle\Phi|u_e\rangle=0$ we get
\Beq
\beta\sum_{\gamma=0} \frac{|a_\gamma|^2}{\mE_\gamma-\mE}=\beta g(\mE)=0,
\Eeq
where $a_\gamma=\langle \psi_\gamma|\Phi\rangle$ and we have defined
\Beq
g(\mE)\equiv\sum_{\gamma=0} \frac{|a_\gamma|^2}{\mE_\gamma-\mE}.
\Eeq
Now either $\beta=0$ or $g(\mE)=0$. If $\beta=0$, then from equation \eqref{eq:umin} we see that  $\min[\mE]=\mEm$ is obtained if  $u_e=\psi_0$: the ground state of $H_2$, which is radially symmetric and has no nodes. This contradicts $\langle \Phi|u_e\rangle=0 $. Hence $\beta\ne0$ and we have $g(\mE)=0$. We need to find the smallest root of $g(\mE)=0$

We will now make use of the technical properties of $H_2$ specified at the beginning of this appendix to analyze the minimum value of $\mE$ that satisfies $g(\mE)=0$. For the lowest radially symmetric eigenstate (ground state of $H_2$ without any nodes), $a_0=\langle \psi_0|\Phi\rangle\ne0$. Moreover, from (b), $\mE_0<0$. For any radially asymmetric (angular) eigenstate $a_\gamma=0$ since that eigenstate will be orthogonal to $\Phi$. In particular if $\psi_1$ is the lowest angular eigenstate, then $a_1=\langle \psi_1|\Phi\rangle=0$ and by our assumption (b), has $\mE_1=0$. If $\psi_2$ is the next radial eigenstate, then by (b), $\mE_2>0$. Now consider the behavior of $g(\mE)$ for $\mE<\mE_0$. In this domain, $g(\mE)>0$. For $\mE_0<\mE<\mE_2$, $g(\mE)$ varies monotonically from $-\infty$ to $+\infty$ and crosses $0$ for the `first' time. Hence if $g(\mE)=0$ in this domain, the root $\mE$ is the smallest root $\mEm$.{\footnote{$g(\mE)$ varies monotonically between every consecutive pair of distinct $\mE_\gamma$.}} Moreover, since $g(\mE)$ varies monotonically from $-\infty$ to $+\infty$ in this domain, the sign of $g(0)$ determines the sign of $\mEm$. Explicitly $g(0)>0\iff \mEm<0$. Hence from \eqref{eq:sign}
\Beq
\label{eq:omegag0}
\textrm{sign}[\min[-\Omega^2]]&=\textrm{sign}[\mEm]=\textrm{sign}[-g(0)].\\
\Eeq
Finally, let us now relate $g(0)$ to $d\langle\Phi|\Phi\rangle/d\alpha$ as follows:
\Beq
&\frac{d}{d\alpha}(H_1\Phi)=0,\\
\Longrightarrow \quad&H_2\frac{d\Phi}{d\alpha}+\Phi=0,\\
\Longrightarrow\quad&\frac{d\Phi}{d\alpha}=-H_2^{-1}\Phi.\\
\Eeq
Multiplying both sides by $\Phi$ and integrating, we get
\Beq
\frac{1}{2}\frac{d\langle\Phi|\Phi\rangle}{d\alpha}
&=-\langle\Phi|H_2^{-1}|\Phi\rangle\\
&=-\sum_{\gamma=0}\frac{|a_\gamma|^2}{\mE_\gamma}\\
&=-g(0).
\Eeq
Thus using the above result and equation \eqref{eq:omegag0}, we finally have
\Beq
\boxed{\textrm{sign}[\min[-\Omega^2]]
=\textrm{sign}\left[\frac{dN}{d\alpha}\right] \quad\quad {\textrm{where}}\quad\quad N\equiv\langle\Phi|\Phi\rangle=\int \Phi^2d^d\trho}
\Eeq

Let us now turn to the justification of the property (a) of $H_1$ assumed in the proof. We will show that $\langle u_e|H_1^{-1}|u_e\rangle$ is positive definite. Note that the eigenvalue problem $H_1\Psi_\beta=E_{\beta}\Psi_{\beta}$ is the time independent Schrodinger equation for a particle of mass $m=1$ in a radial potential well $V_1(\trho)=(\alpha/2)\left[1-(3/4)\textrm{sech}^2(\sqrt{\alpha}\trho)\right]$. Using the profile equation \eqref{eq:profile} we get $H_1\Phi=0$ where $\Phi$ has no nodes. This implies that $\Psi_0=\Phi$ is the unique ground state of $H_1$  (up to a normalization) with energy $E_0=0$ and all other eigenvalues must be greater than $0$. Moreover one has the orthonormal set of excited states $\{\Psi_\beta\}$ with $\beta\ne0$ which satisfies $\langle\Psi_\beta|\Phi\rangle=0$. For any state which belongs to this subspace spanned by $\Psi_\beta$, the operator $H_1$ is positive definite. Hence $H_1^{-1}$ exists on this subspace and is also positive definite. This follows from   $\langle\Psi_\beta|H_1^{-1}|\Psi_\gamma\rangle=E^{-1}_\beta\delta_{\beta\gamma}$. Now, from equation \eqref{eq:eigen}, note that $-\Omega^2\langle\Phi|u_e\rangle=\langle H_1\Phi |H_2u_e\rangle=0$. Hence for $\Omega\ne 0$, $u_e$ lies in the space spanned by $\{\Psi_\beta\}$. Thus $\langle u_e|H_1^{-1}|u_e\rangle$ is positive definite.  We still need to show that $H_2$ has only one bounded eigenmode with a negative eigenvalue, and the lowest angular eigenfunction has $0$ eigenvalue. This is somewhat involved, and we refer the reader to \cite{Vakhitov:1975} where this is discussed further.

\end{document}